\documentclass[aps,preprint]{revtex4}%
\usepackage{amsfonts}
\usepackage{amsmath}
\usepackage{amssymb}
\usepackage{graphicx}%
\begin{document}
\title{The mass formula for an exotic BTZ black hole}
\author{Baocheng Zhang}
\email{zhangbc.zhang@yahoo.com} \affiliation{School of Mathematics
and Physics, China University of Geosciences, Wuhan 430074, China}
\keywords{BTZ black hole; mass formula; Weak Cosmic Censorship}
\pacs{04.20.Cv, 04.70 Bw, 04.20 Dw}

\begin{abstract}
An exotic Ba\~{n}ados-Teitelboim-Zanelli (BTZ) black hole has an
angular momentum larger than its mass in three dimension (3D), which
suggests the possibility that cosmic censorship could be violated if
angular momentum is extracted by the Penrose process. In this paper,
we propose a mass formula for the exotic BTZ black hole and show no
violation of weak cosmic censorship in the gedanken process above by
understanding properly its mass formula. Unlike the other black
holes, the total energy of the exotic BTZ black hole is represented
by the angular momentum instead of the mass, which supports a basic
point of view that the same geometry should be determined by the
same energy in 3D general relativity whose equation of motion can be
given either by normal 3D Einstein gravity or by exotic 3D Einstein
gravity. However, only the mass of the exotic black hole is related
to the thermodynamics and other forms of energy are
\textquotedblleft dumb\textquotedblright, which is consistent with
the earlier thermodynamic analysis about exotic black holes.

\end{abstract}
\maketitle

\section{Introduction}

The two-parameter Ban\~{a}dos, Teitelboim, and Zanelli (BTZ) solution
\cite{btz92} represents a rotating black hole in three-dimensional (3D)
spacetime with a negative cosmological constant. The expressions for the
parameters in terms of conserved quantities of a black hole is
theory-dependent. In the normal 3D Einstein gravity, the identification of the
mass $M$ and the angular momentum $J$ of a black hole gives some similar
behaviors to the Kerr black-hole solution of 4D Einstein equations. In
particular, the mass formula can be gotten directly along the line of the
Smarr's method \cite{ls72} for Kerr black holes; that is%
\begin{equation}
M=\frac{1}{2}TS+\Omega J\label{nsf}%
\end{equation}
where, according to black hole thermodynamics \cite{bch73}, $T$ is the
temperature, $S$ is the entropy, and $\Omega$ is the angular velocity. Like
the Smarr formula for Kerr black holes, the mass $M$ represents the total
energy (in units for which $c=\hbar=1$) which includes the surface energy and
the rotation energy.

In the exotic 3D Einstein gravity \cite{ew88}, however, the parameters have to
be reinterpreted with the reversed roles for mass $M_{E}$ and angular momentum
$J_{E}$, compared with the case of normal Einstein gravity; i.e. $M_{E}\sim
J,J_{E}\sim M$. Usually, such black holes are thought to be \textquotedblleft
exotic\textquotedblright\ which were discovered first in Ref. \cite{cgm95},
and their thermodynamics have been discussed in our recent works
\cite{tz13,bz13}. Coincidentally, the mass formula of an exotic BTZ black hole
has the same form with Eq. (\ref{nsf}),%
\begin{equation}
M_{E}=\frac{1}{2}TS_{E}+\Omega J_{E}\label{esf}%
\end{equation}
where $S_{E}$ is the entropy of the exotic BTZ black hole, which is
proportional to the length of the inner horizon instead of the event horizon
\cite{tz13,sns06,yt07}. To ensure the existence of an event horizon, the
parameters must satisfy a condition%
\begin{equation}
J_{E}\geqslant\ell M_{E}\label{ewc}%
\end{equation}
where $\ell$ is the AdS radius. From this condition, it is seen that
if the mass $M_{E}$ is regarded as the total energy for the exotic
BTZ black hole, it cannot include all the energy that is contributed
by the angular moment. Thus, to investigate the partition of energy
for the exotic BTZ black hole is a natural and significant task for
understanding the thermodynamics better. In fact, the reversed role
for the mass and angular momentum of the exotic BTZ black hole also
caused some other questions that need to be interpreted further. One
is related to the total energy of an exotic BTZ black hole, as
mentioned above. If the total energy is $M_{E}$, the exotic black
hole will have the different energy from the normal one (that is
$M$). But the two kinds of black holes correspond to the same
geometry of spacetime with the same geometric variables, i.e. the
positions of the inner and outer horizon, the temperature, the
angular velocity. In particular, it is noted that the two kinds of
black holes are the solution of 3D Einstein equation which can be
obtained either by 3D normal Einstein gravity or by exotic Einstein
gravity. Thus according to general relativity, only the same energy
leads to the same geometry of spacetime. We conclude that the total
energy must not be $M_{E}$, but what is the total energy of an
exotic BTZ black hole? The other one is related to the condition
(\ref{ewc}). According to Penrose process \cite{pf71}, the energy
associated with the angular momentum $J_{E}$ can be extracted by a
physically feasible process. But such extraction of energy will lead
to violation of the condition (\ref{ewc}) or weak cosmic censorship
(WCC) \cite{rp65}. So how to clarify the question is also a purpose
of this paper.

In this paper, we will investigate the mass formula for exotic BTZ
black holes along the line of Smarr's method, and compare them with
the reversible and irreversible transformations of a black hole made
by Christodoulou \cite{dc70}. The two potential questions will also
be answered in the process of investigation. Actually, the different
recognition of parameters in the exotic BTZ black hole does not
change the total energy which determines the geometry of spacetime
according to Einstein's general relativity. Meanwhile, WCC will be
not violated. In the next section we will first review the
thermodynamics of BTZ black holes for the two cases, based on which
we will study the mass formula for the exotic black holes.

\section{BTZ black holes}

In this section we will give the related information about BTZ black holes
both for the normal and exotic cases, as obtained in our previous works
\cite{tz13,bz13}. Start with the BTZ metric%
\begin{equation}
ds^{2}=-N^{2}dt^{2}+N^{-2}dr^{2}+r^{2}\left(  N^{\phi}dt+d\phi\right)
^{2},\label{btz}%
\end{equation}
where $\phi$ is an angle with the period $2\pi$ as the identification of the
black hole spacetime. The functions $N^{2}$ and $N^{\phi}$ are
\begin{equation}
N^{2}=-8Gm+\frac{r^{2}}{\ell^{2}}+\frac{16G^{2}j^{2}}{r^{2}},N^{\phi}%
=\frac{4Gj}{r^{2}},
\end{equation}
where $G$ the 3D Newton constant. Its Killing horizons are found by setting
$N^{2}=0$; this gives
\begin{equation}
r_{\pm}=\sqrt{2G\ell\left(  \ell m+j\right)  }\pm\sqrt{2G\ell\left(  \ell
m-j\right)  },
\end{equation}
We may assume without loss of generality that $j\geqslant0$ and assume that
$\ell m\geqslant j$, to ensure the existence of an event horizon at $r=r_{+}$.

The BTZ metric can be solved either in the normal Einstein gravity with the
Lagrangian
\begin{equation}
L=\frac{1}{8\pi G}\left(  e_{a}R^{a}-\frac{1}{6\ell^{2}}\epsilon^{abc}%
e_{a}e_{b}e_{c}\right)  ,
\end{equation}
or in the exotic Einstein gravity with the Lagrangian
\begin{equation}
L_{E}=\frac{\ell}{8\pi G}\left[  \omega_{a}\left(  d\omega^{a}+\frac{2}%
{3}\epsilon^{abc}\omega_{b}\omega_{c}\right)  -\frac{1}{\ell^{2}}e_{a}%
T^{a}\right]  ,
\end{equation}
where the Lagrangians are given with 3-forms in which $e^{a}$
($a=0,1,2$) is the dreibein 1-forms, $\omega^{a}$ is Lorentz
connection 1-forms, and their torsion and curvature 2-form field
strengths are $T^{a}=de^{a}+\epsilon
^{abc}\omega_{b}e_{c},R^{a}=d\omega^{a}+\frac{1}{2}\epsilon^{abc}\omega
_{b}\omega_{c}$. It is easily checked that the two Lagrangians give
the same equation of motion which is the 3D Einstein equation, but
their parities are different, which cause some different
thermodynamic interpretations for the two cases. With the BTZ metric
(\ref{btz}), the thermodynamic parameters for the normal BTZ black
holes are
\begin{equation}
M=m,J=j,S=\frac{\pi r_{+}}{2G},
\end{equation}
and the parameters for the exotic ones are%
\begin{equation}
M_{E}=j/\ell,J_{E}=\ell m,S_{E}=\frac{\pi r_{-}}{2G},
\end{equation}
where the exotic forms had been interpreted and the corresponding
thermodynamic laws had been given in our recent work \cite{tz13} and here we
do not elaborate further. In particular, the temperature $T$ and the angular
velocity $\Omega$ take the same values%
\begin{equation}
T=\frac{r_{+}^{2}-r_{-}^{2}}{2\pi r_{+}\ell^{2}},\Omega=\frac{r_{-}}{\ell
r_{+}}%
\end{equation}
for the two cases, which present the geometric properties for the 3D BTZ
spacetime background.

\section{Mass Formula}

\subsection{Normal BTZ black holes}

In this subsection we will revisit the mass formula for a normal BTZ black
hole. It has to be pointed out that the BTZ black hole is asymptotic anti-de
Sitter, different from the Kerr black hole that is asymptotic flat. Generally,
the asymptotic properties will influence the definitions of conserved charges
\cite{ad82,dt02,acz00}, and for the case of 3D gravity see the discussion of
Ref. \cite{gc03}. But we can assume in advance that the mass and angular
momentum of the BTZ black hole has been defined well, i.e. in the normal 3D
Einstein gravity we can identify the mass and angular momentum from the metric
(\ref{btz}) as given in the last section. For exotic BTZ black holes discussed
in the next subsection, we will conform to this assumption.

Start with the mass differential $dM$ according to the thermodynamic first
law,
\begin{equation}
dM=TdS+\Omega dJ.
\end{equation}
Like Smarr's discussion, we choose a path of integration in the space ($S,J$)
to define for a BTZ black hole two energy components: the surface energy
$E_{s}$ by
\begin{equation}
E_{s}=\int_{0}^{S}T\left(  S^{^{\prime}},0\right)  dS^{^{\prime}}; \label{se}%
\end{equation}
and the rotation energy $E_{r}$ by%
\begin{equation}
E_{r}=\int_{0}^{J}\Omega\left(  S,J^{^{\prime}}\right)  dJ^{^{\prime}},\text{
}S\text{ }fixed. \label{re}%
\end{equation}

These integrals can be calculated directly with the variables $M$, $T$, and
$\Omega$ that expressed in terms of $S$ and $J$,%
\begin{align*}
M  &  =\frac{GS^{2}}{2\pi^{2}\ell^{2}}+\frac{\pi^{2}J^{2}}{2GS^{2}},\\
T  &  =\frac{GS}{\pi^{2}\ell^{2}}-\frac{\pi^{2}J^{2}}{GS^{3}},\\
\Omega &  =\frac{\pi^{2}J}{GS^{2}}.
\end{align*}
Then we obtain the results of these integrals as%
\begin{equation}
E_{s}=\frac{GS^{2}}{2\pi^{2}\ell^{2}},E_{r}=\frac{\pi^{2}J^{2}}{2GS^{2}}.
\end{equation}
Thus the total energy is%
\begin{equation}
E=E_{s}+E_{r}=M\text{.} \label{nte}%
\end{equation}

According to Christodoulou \cite{dc70}, the total energy or the mass $M$ of a
BTZ black hole can be divided into an irreducible mass $M_{ir}$ and a
rotational energy $M-M_{ir}$. Now we turn to the calculation of irreducible
mass and see if it is equal to the surface energy $E_{s}$.

Consider a particle of energy $E_{0}$ ($E_{0}\ll M$) sent from infinity into a
3D BTZ black hole and its motion had been analyzed in some previous papers
\cite{fgs93,cmp94}. From its radial geodesics, one knows that the most
efficient Penrose process has to satisfy a condition at the horizon, that is
$\frac{E_{0}}{L}=$ $\frac{4Gj}{r_{+}^{2}}$ where $L$ is the angular momentum
of the particle. Due to the conservation laws of energy and angular momentum,
after the particle drops into the black hole we obtain,%
\begin{equation}
\frac{dM}{dJ}=\frac{J}{\ell\left(  \ell M+\sqrt{\left(  \ell M\right)
^{2}-J^{2}}\right)  }.
\end{equation}
Integrate the relation by taking the initial values $J_{0}=0,M_{0}=M_{ir}$
which is also the starting point of the reversible transformation, and the
irreducible mass is gotten as%
\begin{equation}
M_{ir}=\frac{r_{+}^{2}}{8G\ell^{2}}.
\end{equation}
Using the entropy $S=\frac{\pi r_{+}}{2G}$, we have $M_{ir}=$ $E_{s}$.
Moreover, the relation $\delta M_{ir}\geqslant0$ also implies the second law
of black hole thermodynamics.

\subsection{Exotic BTZ black holes}

Since the first law holds for the exotic black hole, we can use the similar
definitions as Eqs. (\ref{se}) and (\ref{re}) to discuss the energy partition.
At first, we have to work out the expressions of these variables $M_{E}$, $T$
and $\Omega$ in terms of $S_{E},J_{E}$; they are%
\begin{align}
M_{E} &  =\frac{1}{\ell}\left(  \frac{2GS_{E}^{2}J_{E}}{\pi^{2}\ell}%
-\frac{G^{2}S_{E}^{4}}{\pi^{4}\ell^{2}}\right)  ^{\frac{1}{2}};\nonumber\\
T &  =\frac{2GS_{E}}{\ell\pi^{2}}\left(  J_{E}-\frac{GS_{E}^{2}}{\pi^{2}\ell
}\right)  \left(  \frac{2GS_{E}^{2}J_{E}}{\pi^{2}\ell}-\frac{G^{2}S_{E}^{4}%
}{\pi^{4}\ell^{2}}\right)  ^{-\frac{1}{2}};\nonumber\\
\Omega &  =\frac{GS_{E}^{2}}{\pi^{2}\ell}\left(  \frac{2GS_{E}^{2}J_{E}}%
{\pi^{2}\ell}-\frac{G^{2}S_{E}^{4}}{\pi^{4}\ell^{2}}\right)  ^{-\frac{1}{2}%
}.\label{ebhtq}%
\end{align}
where an implicit assumption that $J_{E}\geqslant\frac{GS_{E}^{2}}{2\pi
^{2}\ell}$ has been enforced in order to ensure the real values of these
physical variables.

Thus, one can choose $\frac{GS_{E}^{2}}{2\pi^{2}\ell}$ as the lower limit of
the angular momentum. With Eqs. (\ref{se}) and (\ref{re}), we get the surface
energy of the exotic BTZ black hole as $E_{s}=0$ and the rotation energy
$E_{r}=M_{E}$, which is consistent with the relation $M_{E}=E_{s}+E_{r}$. But
this takes the ground state with $M_{E0}=0$ and $J_{E0}=\frac{GS_{E}^{2}}%
{2\pi^{2}\ell}$, which is inconsistent with the former assumption
that $M_{E}$ represents the total energy in the mass formula. Thus,
either the total energy is not given by $M_{E}$ or the analysis here
is unfeasible. Since the thermodynamics about exotic BTZ black holes
had been confirmed \cite{tz13,bz13}, we guess that the analysis is
feasible but some changes have to be made.

Define a new reasonable ground state with $\ell M_{E0}=J_{E0}$ which is also
the extreme state of the exotic BTZ black hole. With such ground state, we
have to take the constraint further as $J_{E}\geqslant J_{C}=\frac{GS_{E}^{2}%
}{\pi^{2}\ell}$ by the expression for mass $M_{E}$. Thus $J_{C}$ should be
taken as the lower limit, which will be explained further below when
discussing the irreducible mass. Now we calculate the surface energy%
\begin{equation}
E_{Es}=\int_{0}^{S_{E}}T\left(  S_{E}^{^{\prime}},J_{C}\right)  dS_{E}%
^{^{\prime}}=\frac{GS_{E}^{2}}{\pi^{2}\ell^{2}};
\end{equation}
which is just the energy associated with the part of the angular moment
$J_{C}$, or the energy of ground state, and the rotation energy
\begin{equation}
E_{Er}=\int_{J_{C}}^{J_{E}}\Omega\left(  S_{E},J_{E}^{^{\prime}}\right)
dJ_{E}^{^{\prime}}=\frac{1}{\ell}\left(  \frac{2GS_{E}^{2}J_{E}}{\pi^{2}\ell
}-\frac{G^{2}S_{E}^{4}}{\pi^{4}\ell^{2}}\right)  ^{\frac{1}{2}}-\frac
{GS_{E}^{2}}{\pi^{2}\ell^{2}}. \label{ern}%
\end{equation}
Instantly, one obtain $M_{E}=E_{Es}+E_{Er}$. From the calculation of $E_{Er}$,
one might think that the rotation energy $E_{Er}$ is just the energy
associated with the part with the range of angular momentum from $J_{C}$ to
$J_{E}$ (we will call it the part of $J_{E}-J_{C}$), but a further analysis
finds that $E_{Er}\leqslant$ energy (associated with the part of $J_{E}-J_{C}%
$), i.e. $\ell E_{Er}=$ $\sqrt{\frac{2GS_{E}^{2}J_{E}}{\pi^{2}\ell}%
-\frac{G^{2}S_{E}^{4}}{\pi^{4}\ell^{2}}}-\frac{GS_{E}^{2}}{\pi^{2}\ell^{2}%
}=\sqrt{J_{E}^{2}-\left(  J_{E}-\frac{GS_{E}^{2}}{\pi^{2}\ell}\right)  ^{2}%
}-J_{C}\leqslant J_{E}-J_{C}$. So the energy $E_{Er}$ does not
include all the energy contributed by the angular momentum
$J_{E}-J_{C}$. An energy contributed by the part of angular moment
$J_{E}-J_{C}-\ell E_{Er}$ has to be introduced for the energy
conservation, that is, $E_{ER}=\frac{1}{\ell}\left(
J_{E}-J_{C}\right) -E_{Er}=\frac{J_{E}}{\ell}-M_{E}$. Then the total
energy for an exotic BTZ black hole described by the metric
(\ref{btz}) can be
written as%
\begin{equation}
E_{E}=E_{Es}+E_{Er}+E_{ER}=\frac{J_{E}}{\ell}\text{.} \label{ete}%
\end{equation}

Comparing Eq. (\ref{nte}) with Eq. (\ref{ete}), we find that $E=E_{E}$ which
shows that the total energies for two different black holes are the same,
which determine the same geometry of spacetime. In order to confirm this
analysis, we will discuss the reversible and irreversible transformations for
an exotic BTZ black hole. Using the same process of a particle falling into
the black hole, it is given an equation as%
\begin{equation}
\frac{dM_{E}}{dJ_{E}}=\frac{M_{E}}{\ell\left(  J_{E}+\sqrt{J_{E}^{2}-\left(
\ell M_{E}\right)  ^{2}}\right)  },
\end{equation}
where the equation is also derived from the relation
$\frac{E_{0}}{L}=$ $\frac{4Gj}{r_{+}^{2}}$ which is obtained from
the geodesic motion of particles \cite{fgs93,cmp94} and independent
on the concrete 3D gravity models, but the corresponding parameters
have to be matched with the exotic black holes. Integrating the
equation but taking the initial values as
$\frac{J_{E0}}{\ell}=M_{E0}=M_{E-ir}$, we obtain the irreducible mass as%
\[
M_{E-ir}=\frac{r_{-}^{2}}{4G\ell^{2}},
\]
which is equal to $E_{Es}$ by using the entropy $S_{E}=\frac{\pi r_{-}}{2G}$.
Actually by the equality $M_{E-ir}=E_{Es}$, one can also obtain the expression
of the entropy as $S_{E}=\frac{\pi r_{-}}{2G}$ for the exotic BTZ black hole.
Again the relation $\delta M_{E-ir}\geqslant0$ supports the second law of
black hole thermodynamics for an exotic BTZ black hole, as expected. Moreover,
it is noted that $M_{E-ir}=J_{C}/\ell$, which gives the angular momentum of
ground state as $J_{E0}/\ell=M_{E-ir}$. When we put $J_{E}/\ell=J_{E0}%
/\ell=M_{E-ir}$ into the expression of $M_{E}\left(  S_{E},J_{E}\right)  $, we
get $M_{E}=M_{E-ir}=M_{E0}$, which indicates that the ground state with $\ell
M_{E0}=J_{E0}$ are physically reasonable and self-consistent. Then can we take
the values with such relation $\frac{J_{E0}}{\ell}<M_{E0}$? The answer is
negative because such initial values will lead to an imaginary irreducible mass.

It is noted that $M_{E}-M_{E-ir}$ cannot be considered as all the
rotation energy, since
$M_{E}-M_{E-ir}\leqslant\frac{J_{E}}{\ell}-M_{E-ir}$. So we define
the excess $\left(  \frac{J_{E}}{\ell}-M_{E-ir}\right)  -\left(
M_{E}-M_{E-ir}\right)  =\frac{J_{E}}{\ell}-M_{E}$ as a part of the
total energy, which is just $E_{ER}$ mentioned before. Again we show
that the total energy of an exotic BTZ black hole is $J_{E}$ which
consists of three parts: the surface energy, the rotation energy,
and the redundant rotation energy. Here we make some discussions for
these energies. Firstly, different from the general understanding
for the surface energy, it consists of the rotation energy
associated with the part of angular momentum $J_{C}$. In particular,
the rotation energy included in the surface energy cannot be
extracted through the Penrose process, which ensure the validity of
WCC. Secondly, the rotation energy is the same as usual, and can be
extracted through the Penrose process. Finally, the redundant
rotation energy is interesting, since it can be extracted through
the Penrose process but cannot be obtained by the standard
calculation for the rotation energy, i.e. by Eq. (\ref{ern}). And
its existence looks more like a kind of indication that WCC is
preserved (i.e. only if $E_{ER}\geqslant0$, WCC is not violated).
Then through a brief process, these energies can be understood
further. Given an exotic BTZ black hole with
$M_{E}\leqslant\frac{J_{E}}{\ell}$, one can extract the rotation
energy $M_{E}-M_{E-ir}$ through Penrose process. Then the energy of
the black hole is left as $\frac{J_{E}}{\ell}-\left(
M_{E}-M_{E-ir}\right)  =$ $\left( \frac{J_{E}}{\ell}-M_{E}\right) +$
$M_{E-ir}=$ $E_{ER}+M_{E-ir}$. If the energy $E_{ER}$ is taken out,
the extreme black hole appears, but according to the third law of
black hole thermodynamics, the energy $E_{ER}$ cannot extracted
completely during a finite time. So it is not necessary to worry
about violation of WCC due to the excessive extraction of rotation
energy.

Then, we want to investigate the extreme situation. For a normal BTZ black
hole, its extreme condition is $\ell M=J$, for which $E_{s}=E_{r}=\frac{M}{2}%
$. The non-zero value of the rotation energy indicates that even in the
extreme situation, the rotation energy can still be extracted through Penrose
process. This will not cause any problems for the normal BTZ black hole, but
at the extreme situation, the extraction of the rotation energy will lead to
violation of WCC for an exotic BTZ black hole. Fortunately, under the extreme
condition $\ell M_{E}=J_{E}$ of the exotic BTZ black hole, the energy that can
be extracted through the Penrose process has been exhausted, i.e.
$E_{Er}=E_{ER}=0$, and the extra rotation energy that appears with the form of
the surface energy or the irreducible mass cannot be reduced further and thus
protects the WCC from violation by the extraction of rotation energy. In
particular, due to the exotic components of the surface energy, the scenario
of particle with the angular momentum infalling into a black hole will not
lead to any violation of WCC, that can refer to the discussion in Ref.
\cite{rc11}.

A further comment contributes to such a question: since the total energy is
$E_{E}$ that is not equal to $M_{E}$, it seems better to work out the
differential form of energy $E_{E}$ to make the black hole behave like a
thermal system satisfying the first law of thermodynamics. From the discussion
above, we have
\begin{equation}
dE_{E}=dE_{Es}+dE_{Er}+dE_{ER},
\end{equation}
where $dE_{Es}=0$ since the energy associated with $E_{Es}$
represents its irreducible mass. Using the expression for $E_{ER}$,
the first law becomes $dM_{E}=dE_{Er}.$Then using Eqs. (\ref{ebhtq})
and (\ref{ern}) and noticing that $dJ_{C}=0$, a straight calculation
gives $dE_{Er}=TdS_{E}+\Omega dJ_{E}$. Thus we get
\begin{equation}
dM_{E}=TdS_{E}+\Omega dJ_{E},
\end{equation}
which is consistent with our earlier results for the first law of
thermodynamics for an exotic BTZ black hole \cite{tz13}. In
particular, this interpretation also applies to the exotic BTZ
black-hole solution obtained in \textquotedblleft
BCEA\textquotedblright\ theory \cite{cgm95,cg91} which minimally
couples topological matter to 3D Einstein gravity without the
cosmological constant if one associates this term $dE_{ER}$ with the
exotic topological matter. Moreover, such interpretation is very
like that in the case of charged rotating BTZ black holes
\cite{as07}, in which the total energy is not represented by the
mass, but the expression of the first law there includes an extra
term $P_{r}dA$ that is from the contribution of electromagnetic
matter.

Finally, the general Hamiltonian analysis gives the energy of the
exotic BTZ black hole as $M_{E}$ which is the on-shell values of the
asymptotic generators for time translations \cite{bcv13}, but it has
to be pointed out that such energy is only related to the dynamic or
thermodynamic process. Whether there are other forms of energy that
are not included in the thermodynamic process is unclear for the
Hamiltonian form. For our case, a simple relation (\ref{ewc}) tells
us that the mass $M_{E}$ cannot represent the whole energy and it
must require some other forms of energy. Based on our analysis,
these forms of energy are the surface energy $E_{Es}$ which ensures
that the initial state is consistent with WCC and the redundant
rotation energy $E_{ER}$ which ensures that the final state will not
violate WCC. Since the extra forms of energy are \textquotedblleft
dumb\textquotedblright\ which do not contribute to the thermodynamic
process, the analysis here is consistent with the earlier
understanding for the thermodynamics about exotic BTZ black holes.

\section{Conclusion}

In this paper, we have investigated the mass formulas for the normal
and exotic BTZ black holes and have presented their differences. For
normal BTZ black holes, the understanding of the mass formula is
nearly the same with that for Kerr black holes. For exotic BTZ black
holes, however, the understanding of the mass formula is far from
the usual situation. Firstly, the total energy is represented by the
angular momentum, instead of the mass. Then the surface energy that
can be interpreted as irreducible mass includes one part of the
rotation energy that cannot be extracted by the Penross process.
This ensures the validity of WCC. We have also showed that the total
energy will not change, although the parameters ($m,j$) are read by
the different conserved quantities of a black hole in the normal and
exotic 3D gravity models. This ensures that the same geometry of
spacetime is described by the metric (\ref{btz}) in the two
different models of 3D general relativity. Thus the discussion of
the mass formula in this paper improves further the understanding
about thermodynamics of exotic BTZ black holes.

Further, we pointed out that the form presented in Eq. (\ref{nsf})
can be applicable to all BTZ black holes in different 3D gravity
models such as topological massive gravity \cite{djt82}, new massive
gravity \cite{bht09}, general massive gravity \cite{bht09,hrtz12}
and so on. The difference between these modified theories and the
normal Einstein gravity is different from the difference between the
exotic Einstein gravity and the normal Einstein gravity. For the
former, they have different equations of motion, but for the latter,
the equation of motion is the same one which is 3D Einstein
equation. So in this paper, we discuss only the case for 3D general
relativity which should inherit the basic point of view in 4D
general relativity that says the geometry is determined by the total
energy. But whether the modified gravity theories could also inherit
such point of view should deserve further investigation. Moreover,
it has to be stressed that the universal form (\ref{nsf}) of mass
formula is not the 3D extension of Smarr formula which requires an
extended thermodynamic phase space in a new method called as
\textquotedblleft black hole chemistry\textquotedblright\
\cite{cm95}, as pointed out in a recent Ref. \cite{fmm15} that the
scaling properties of the various thermodynamic parameters for BTZ
black holes in Eq. (\ref{nsf}) are inconsistent with the requirement
of Smarr formula. But whether the extended Smarr formula in 3D can
be consistently applied to all cases as done for the universal form
of Eq. (\ref{nsf}) is unclear now and deserve a further
investigation.

\section{Acknowledgements}

The author would like to thank Prof. Townsend for bringing the
problem to him and for reading and revising this paper, and thanks
the anonymous referee for his/her critical comments and helpful
advice. This work is supported by Grant No. 11374330 of the National
Natural Science Foundation of China, the Open Research Fund Program
of the State Key Laboratory of Low-Dimensional Quantum Physics in
Tsinghua University, and the Fundamental Research Funds for the
Central Universities, China University of Geosciences (Wuhan)(No.
CUG150630).

\end{document}